\documentclass[%
 reprint,
 amsmath,amssymb,
 aps,
]{revtex4-1}
\usepackage{graphicx}% Include figure files
\usepackage{dcolumn}% Align table columns on decimal point
\usepackage{bm}% bold math
\usepackage{amsmath,amssymb}
\usepackage{slashed}
\usepackage{graphicx}
\usepackage{dsfont}
\usepackage{subfig}
\usepackage{bm}
\usepackage[top=1.0in,bottom=1.0in,left=1.0in,right=1.0in]{geometry}
\usepackage[colorlinks,linkcolor=blue,citecolor=blue]{hyperref}
\usepackage{CJKutf8}
\usepackage{tikz-feynhand}

\begin{document}
\setlength{\oddsidemargin}{0.5cm}
\setlength{\topmargin}{-0.1cm}
\setlength{\textheight}{21cm}
\setlength{\textwidth}{15cm}
\newcommand{\be}{\begin{equation}}
\newcommand{\ee}{\end{equation}}
\newcommand{\bea}{\begin{eqnarray}}
\newcommand{\eea}{\end{eqnarray}}
\newcommand{\ba}{\begin{eqnarray}}
\newcommand{\ea}{\end{eqnarray}}

\newcommand{\fslash}{\hspace{-1.4ex}/\hspace{0.6ex} }
\newcommand{\Dslash}{D\hspace{-1.6ex}/\hspace{0.6ex} }
\newcommand{\Wslash}{W\hspace{-1.6ex}/\hspace{0.6ex} }
\newcommand{\pslash}{p\hspace{-1.ex}/\hspace{0.6ex} }
\newcommand{\kslash}{k\hspace{-1.ex}/\hspace{0.6ex} }
\newcommand{\underkslash}{{\underline k}\hspace{-1.ex}/\hspace{0.6ex} }
\newcommand{\epslash}{{\epsilon\hspace{-1.ex}/\hspace{0.6ex}}}
\newcommand{\partslash}{\partial\hspace{-1.6ex}/\hspace{0.6ex} }

\newcommand{\nn}{\nonumber}
\newcommand{\Tr}{\mbox{Tr}\;}
\newcommand{\tr}{\mbox{tr}\;}
\newcommand{\ket}[1]{\left|#1\right\rangle}
\newcommand{\bra}[1]{\left\langle#1\right|}
\newcommand{\rhoraket}[3]{\langle#1|#2|#3\rangle}
\newcommand{\brkt}[2]{\langle#1|#2\rangle}
\newcommand{\pdif}[2]{\frac{\partial #1}{\partial #2}}
\newcommand{\pndif}[3]{\frac{\partial^#1 #2}{\partial #3^#1}}
\newcommand{\pbm}[1]{\protect{\bm{#1}}}
\newcommand{\avg}[1]{\left\langle #1\right\rangle}
\newcommand{\vnabla}{\mathbf{\nabla}}
\newcommand{\notes}[1]{\fbox{\parbox{\columnwidth}{#1}}}
\newcommand{\pair}{\raisebox{-7pt}{\includegraphics[height=20pt]{pair0.pdf}}}
\newcommand{\paircrs}{\raisebox{-7pt}{\includegraphics[height=20pt]{pair0cross.pdf}}}
\newcommand{\paircc}{\raisebox{-7pt}{\includegraphics[height=20pt]{pair0cc.pdf}}}
\newcommand{\paircrscc}{\raisebox{-7pt}{\includegraphics[height=20pt]{pair0crosscc.pdf}}}
\newcommand{\pairloop}{\raisebox{-7pt}{\includegraphics[height=20pt]{pairloop.pdf}}}
\newcommand{\pairloopf}{\raisebox{-7pt}{\includegraphics[height=20pt]{pairloop4.pdf}}}
\newcommand{\pairlooph}{\raisebox{-7pt}{\includegraphics[height=20pt]{pair2looph.pdf}}}

%\preprint{APS/123-QED}

\title{Pion gravitational form factors in the QCD instanton vacuum I}

\author{Wei-Yang Liu}
\email{wei-yang.liu@stonybrook.edu}
\affiliation{Center for Nuclear Theory, Department of Physics and Astronomy, Stony Brook University, Stony Brook, New York 11794-3800, USA}

\author{Edward Shuryak}
\email{edward.shuryak@stonybrook.edu}
\affiliation{Center for Nuclear Theory, Department of Physics and Astronomy, Stony Brook University, Stony Brook, New York 11794-3800, USA}

\author{Christian Weiss}%
 \email{weiss@jlab.org}
\affiliation{Theory Center, Jefferson Lab, Newport News, VA 23606, USA}

\author{Ismail Zahed }
\email{ismail.zahed@stonybrook.edu}
\affiliation{Center for Nuclear Theory, Department of Physics and Astronomy, Stony Brook University, Stony Brook, New York 11794--3800, USA}

\date{\today}% It is always \today, today,
\begin{abstract}
The pion form factors of the QCD energy-momentum tensor (EMT) are studied in the instanton liquid model (ILM) of the QCD vacuum. In this approach the breaking of conformal symmetry is encoded in the form of stronger-than-Poisson fluctuations in the number of instantons.
For the trace of the EMT, it is shown that the gluonic trace anomaly term contributes half the pion mass, with the other half coming from the quark-mass-dependent sigma term. The $Q^2$ dependence of the form factors is governed by glueball and scalar meson exchanges.
For the traceless EMT, the spin-0 and 2 form factors are computed at next-to-leading order in the instanton density using effective quark operators. Relations between the gluon and quark contributions to the EMT form factors are derived. The form factors are also expressed in terms of the pion light-front wave functions in the ILM. The results at the low resolution scale of the inverse instanton size are evolved to higher scales using the renormalization group equation.
The ILM results compare well with those of recent lattice QCD calculations.
%The breaking of conformal symmetry in the instanton liquid model (ILM)  is encoded in the form of stronger than Poisson fluctuations  in the number of topological charges, which are shown to contribute half the pion mass, with the other half following from the pion sigma term. In the ILM, the glue contribution to the pion gravitational form factors (GFFs)  from single and pair of instantons, is described by effective scalar and tensor quark operators, at next-to-leading order in the instanton density. We use these results to  derive explicit relations between the gluon and quark contributions to the pion GFF,  at the low resolution fixed by the  inverse instanton size. The results are evolved to larger resolution, using  QCD evolution for the gravitational charges.  The pion GFFs results compare well with those recently reported by lattice collaborations.
\end{abstract}

\maketitle

\section{Introduction}
\label{INTRODUCTION}
\begin{figure}[t]
	\begin{center}
		\includegraphics[width=6cm]{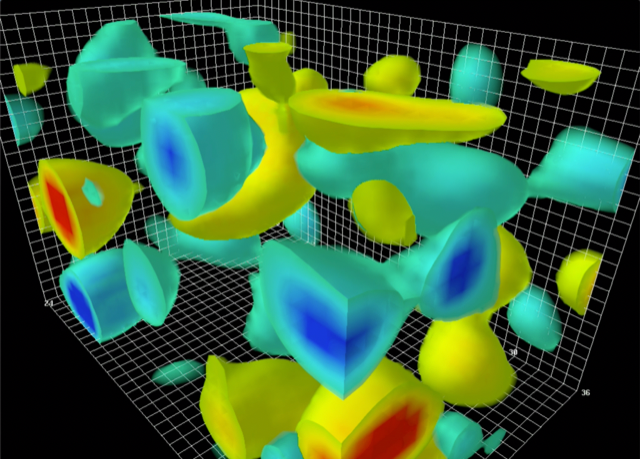}
		\caption{Instantons (yellow) and anti-instantons (blue) configurations in the cooled gluodynamics vacuum~\cite{Leinweber:1999cw}.}
		\label{fig_VAC}
	\end{center}
\end{figure}

The QCD vacuum breaks conformal symmetry,
a mechanism at the origin of most hadronic masses. The nature of the gauge fields at the origin of this breaking is still debated, but there is increasing evidence that at low resolution the QCD
vacuum is populated by instantons and anti-instantons (pseudoparticles), tunneling configurations
between vacua with different topological charges. Gradient flow (cooling) 
techniques have revealed a landscape as illustrated in
Fig.~\ref{fig_VAC}~\cite{Leinweber:1999cw}.  These pseudoparticles are very effective at breaking spontaneously chiral symmetry through fermionic zero modes with fixed chirality (left or right). 

The key features of this landscape are~\cite{Shuryak:1981ff}
\begin{equation}
n_{I+A}\equiv \frac 1{R^4}\approx \frac 1{ {\rm fm}^{4}} \qquad\qquad\frac{\bar \rho}R \approx  \frac 13   \label{eqn_ILM}
\end{equation}
for the instanton plus anti-instanton density and size, respectively.
The hadronic scale $R=1\,{\rm fm}$
emerges as the mean quantum tunneling rate of the quasiparticles. The quasiparticles are sparsely distributed in the 4-dimensional Euclidean space, with a packing fraction 
$$\kappa\equiv \pi^2\bar\rho^4 n_{I+A}\approx 0.1$$
which is key in organising the many-body problem.

In the course of the cooling procedure leading to the landscape of Fig.~\ref{fig_VAC},
hadronic correlation functions undergo only moderate changes from their original lattice 
values, and mostly at short distances; see Ref.~\cite{Schafer:1996wv} and references therein.
The deeply cooled correlation functions, which can be computed using semi-classical methods, 
are therefore generally good approximations to the QCD correlation functions at hadronic distances.
%In the  deeply cooled lattice landscape, most hadronic correlations undergo moderate change mostly at short distances
%~\cite{Schafer:1996wv} (and references therein). 
For completeness, we note that recently it was argued that certain tunneling trajectories beyond
single instantons/antiinstantons, so-called failed instanton-anti-instanton tunnelings or molecules, 
increase the packing fraction from $0.1\rightarrow 0.7$~\cite{Shuryak:2021fsu} and may account for part of the confining potential in the zero cooling approximation. However, this direction will not be pursued here.

The purpose of this work is to derive the energy-momentum tensor (EMT) form factors of the pion in the instanton liquid model (ILM) description of the deeply cooled QCD vacuum. 
The EMT form factors
are of fundamental interest for understanding hadron structure and large-distance dynamics.
The trace of the EMT describes the composition of hadron masses and quantifies the role of 
conformal and chiral symmetry breaking. The traceless part of the EMT describes the cumulative 
twist-2 quark-gluon structure of hadrons and can be interpreted in terms of ``mechanical properties'' 
of the dynamical system \cite{Polyakov:2018zvc}. The pion is of special interest as the Goldstone boson of spontaneously broken chiral symmetry. Constraints on the pion EMT form factors at low momentum transfers $\mathcal{O}(m_\pi^2)$
can be derived using model-independent methods (soft-pion theorems, chiral EFT).
The EMT form factors of the pion have also been analyzed on the lattice~\cite{Hackett:2023nkr,wang2024trace}, 
and using phenomenological models~\cite{Broniowski_2008,Frederico_2009,Masjuan_2013,Fanelli_2016,Freese_2019,Krutov_2021,Raya:2021zrz,Xu:2023izo,Li_2024,Broniowski:2024oyk}. 

The outline of the paper is as follows: in
section~\ref{VACUUM} we briefly review the 
salient features of the instanton vacuum, 
with emphasis on the breaking of conformal symmetry. 
In section~\ref{ANOMALY} we show how the forward matrix elements of the trace of the pion EMT are constrained by the instanton vacuum compressibility,
leading to a pion mass identity. In section~\ref{PIONEMT} 
we first outline the general structure of the pion GFFs
and the threshold constraints from chiral symmetry.
We then compute the form factors of the trace and traceless parts of the EMT in the
instanton vacuum. The trace part is obtained as a combination of a scalar glueball contribution and the pion scalar form factor. 
The traceless part is evaluated at leading and next-to-leading order in the instanton density. We also connect the invariant form factors with the light front 
components of the EMT and express them in terms of the pion light-front wave functions.
In section~\ref{RESULTS} our results are compared to the recently reported lattice simulations for the pion $A$ and $D$ and  gravitational form factors, along with the pertinent radii. Our conclusions are in section~\ref{CONCLUSION}. The pion light front wavefunction and scalar form factors are discussed in the appendices.

\section{QCD instanton vacuum}
\label{VACUUM}
 The size distribution of the instantons
and anti-instantons 
%density 
(their tunneling rate) in the vacuum is well captured semi-empirically by
\begin{equation}
\label{dn_dist}
n(\rho) \sim  {1 \over \rho^5}\big(\rho \Lambda_{QCD} \big)^{b} \, e^{-C\rho^2/R^2}
\end{equation}
with $b=11N_c/3-2N_f/3$ (one loop).  
The small-size distribution follows from the conformal nature of the
instanton moduli and perturbation theory. The large-size distribution is non-perturbative. A variational analysis of the
instanton ensemble including binary interactions \cite{Diakonov:1983hh,Diakonov:1995ea} shows that large-size instantons are suppressed self-consistently by their density as in~(\ref{dn_dist}),
 with $C$ a number of order unity that depends on the specific model of the interactions.

%\subsection{Quantum conformal symmetry breaking and the trace anomaly}

The quantum breaking of conformal symmetry is best captured by the trace of the EMT. 
More specifically, the symmetric and conserved EMT is
\bea
\label{1}
T^{\mu\nu}
=-F^{a\mu\lambda}F^{a\nu}_\lambda+\frac 14 g^{\mu\nu}F^2+ \overline\psi \gamma^{(\mu} i\overleftrightarrow D^{\nu)}\psi\nonumber ,\\
\eea
where $\overleftrightarrow{D}=\frac{1}{2}(\overrightarrow{D}-\overleftarrow{D})$ and $()$ denotes symmetrization. The equation of motion $(i\slashed{D}-m)\psi=0$ has been  applied in Eq.\eqref{1}. The EMT traces to zero classically, but is anomalous quantum mechanically owing 
to its sensitivity to short distance fluctuations, namely
\be
\label{3X}
T^\mu{}_\mu=\frac{\beta(g^2)}{4g^4}F^a_{\mu\nu}F^{a\mu\nu}+m\overline\psi\psi
\ee
after proper regularization,
with  the Gell-Mann-Low beta-function (2 loops)
\be
\label{beta}
\beta(g^2)=-\frac{bg^4}{8\pi^2}-\frac{\bar bg^6}{2(8\pi^2)^2}+{\cal O}(g^8).
\ee
Throughout, we absorb $gF\rightarrow F$ for all gauge field operators; with this normalization
 $$\frac {F^2}{32\pi^2}\rightarrow \frac{(N_++N_-)}{V_4}$$ counts the number of quasiparticles in the 4-volume. In the vacuum with zero vacuum angle, the number of instantons and antiintantons is the same on average, $N_\pm/V_4=\bar N/2V_4$. In leading order in the packing fraction
\be
\label{SCALE}
\left<T^\mu{}_\mu\right>\approx 
\frac {\langle F^2\rangle}{32\pi^2}
\approx-b \bigg(\frac{\bar N}{V_4}\bigg)
%+m\left<\overline \psi\psi\right>
\approx -10\,{\rm fm}^{-4},
\ee
which represents 4 times the vacuum energy density because of the O(4) symmetry in the Euclidean metric. 
The subtraction of the perturbative and divergent contribution in $\langle F^2\rangle$ is assumed through the use of the cooling procedure,
for fixed topological susceptibility.
The scale of the spontaneous breaking of chiral symmetry is also fixed by the finite instanton density, but the contribution of the chiral condensate to the vacuum scalar density  is small since  $mR \approx (8\,{\rm MeV})(1\,{\rm fm})\approx 1/25$.

Some of the quantum  scale fluctuations in QCD are captured in the instanton vacuum
using the grand-canonical description, where the quasiparticle number $N=N_++N_-$ is allowed to fluctuate with the measure
\cite{Diakonov:1995qy,Kacir:1996qn,Nowak:1996aj}
\be
\label{dist}
\mathbb P(N)=e^{\frac{bN}4 }\bigg(\frac {\bar{N}}{N}\bigg)^{\frac {bN}4 }
\ee
to reproduce the vacuum topological compressibility
\be
\label{COMP}
\frac{\sigma_T}{\bar{N}} =\frac{\langle(N-\bar N)^2\rangle}{\bar N}=\frac 4b \approx \frac 1{N_c}
\ee
in agreement with low-energy theorems~\cite{Novikov:1981xi}.
In this formulation the cooled QCD vacuum is a quantum liquid of quasiparticles, with a topological compressibility of about $\frac 13$ at $N_c = 3$, and 
incompressible at large $N_c$.

\section{Pion mass identity}
~\label{ANOMALY}
In a  pion state $|p\rangle$ with the standard normalization 
$\langle p|p^\prime\rangle = 2E(\bm{p})(2\pi)^3 \delta^3(\bm{p}-\bm{p}^\prime)$ with $E(\bm{p})=\sqrt{m_\pi^2+|\bm{p}|^2}$, the matrix element of the EMT is 
\begin{align}\label{forwardT}
\langle p|T^{\mu\nu}|p\rangle=2p^{\mu}p^{\nu} \ .
\end{align}
This form follows from Lorentz invariance and the fact that the matrix element 
of the $T^{00}$ component is the total energy of the state. The matrix element 
of the trace is 
\begin{align}
&\langle p|T^\mu{}_\mu|p\rangle = 2m_\pi^2 ,
\end{align}
which is independent of the reference frame.
With the explicit form of the operator given by the anomaly (\ref{3X}), one obtains
an expression for the pion mass as
\begin{align}
\label{Trace}
m_\pi &= \frac{\langle p|T^\mu{}_\mu|p\rangle}{2 m_\pi} 
\nonumber \\
&= -\frac{b}{32\pi^2} \frac{\langle p| F^2 | p \rangle}{2 m_\pi}
+ \frac{\langle p | m\bar \psi\psi |p\rangle}{2 m_\pi} . 
\end{align}
%\label{Trace}
%\begin{align}
%\label{Trace}
%&\langle p|T^\mu{}_\mu|p\rangle=\nonumber\\
%&\langle p|\bigg(-\frac{b}{32\pi^2} F^2+m\bar \psi\psi\bigg)|p\rangle=2m_\pi^2 . 
%\nonumber\\
%\end{align}
%with  $g^2 F^2\rightarrow F^2$  subsumed. 
The identity (\ref{Trace}) shows that the pion mass is the sum of two terms:
the change of the conformal anomaly or gluon field in the pion state, and the 
average of the scalar quark density. The pion state forms only if chiral 
symmetry is spontaneously broken. Both terms in the mass decomposition vanish
in the chiral limit. We now show that Eq.~(\ref{Trace}) is satisfied in the
ILM and determine the relative magnitude of the two contributions.

For a pion state with zero 3-momentum, denoted by $|\pi \rangle$, 
the gluon contribution to the matrix element in (\ref{Trace}) is obtained from the normalized connected 3-point function as
\begin{align}
\label{corr}
&\frac{\langle \pi | F^2(0) |\pi\rangle}{\langle \pi|\pi\rangle}
\nonumber \\
&={\lim_{T\to\infty}}\frac{\left<J_\pi^\dagger (T/2) F^2(0) J_\pi(-T/2)\right>_{\rm conn}}{\left<J_\pi^\dagger (T/2)J_\pi(-T/2)\right>}.
\end{align}
We use a shorthand notation where the normalization $\langle \pi | \pi \rangle = 2 m_\pi (2\pi)^3 \delta^3 (\bm{0}) = 2 m_\pi V_3$ contains the 3-volume. 
Here $J_\pi (\mp T/2)$ are pion sources projected on zero 3-momentum, at Euclidean times $\mp T/2$. 
The correlation functions are the connected averages over the grand canonical ensemble, including the average over the instanton number fluctuations (\ref{dist}).
Because the gluon operator is evaluated in a state with zero momentum, its position
in the correlation function can be averaged over the spatial coordinate and over the
Euclidean time in the interval between $-T/2$ and $T/2$,
\begin{align}
F^2(0) \rightarrow 
\frac{1}{T V_3} \int_{-T/2}^{T/2} dx_4
\int d^3 x \, F^2(x).
\end{align}
The value of the integrated operator is given by the number of instantons in the volume $V_3 \times T$. The average value of the operator, given by the average number of instantons in the volume, does not contribute to the connected correlator; it cancels when subtracting the vacuum contribution.
The leading non-vanishing contribution to the connected correlator arises from the fluctuations of the number of instantons in the volume described by the distribution Eq.~(\ref{dist})
 \cite{Diakonov:1995qy,Kacir:1996qn}. The relative fluctuations in the volume $V_3 \times T$ are the same as those in the full Euclidean volume $V_4$ and have the variance given by
 Eq.~(\ref{COMP}). We obtain
\begin{widetext}
\begin{align}
& \frac{\left< J_\pi^\dagger (T/2) \, F^2(0) \, J_\pi(-T/2) \right>_{\rm conn}}
{\left<J_\pi^\dagger (T/2)J_\pi(-T/2)\right>}
\nonumber \\
&= \frac{32\pi^2}{V_3 T} \frac{\langle (N - \bar N)^2 \rangle}{\bar N} \,  
\frac{\left. \displaystyle N\frac{d}{dN} \left< J_\pi^\dagger (T/2) J_\pi(-T/2) \right>_N
\right|_{N = \bar N}}
{\left<J_\pi^\dagger (T/2)J_\pi(-T/2)\right>_{\bar N}}
\\
&= \frac{32\pi^2}{V_3 T} \; \frac{4}{b} \;  \left. N\frac{d}{dN}  
\log \left<J_\pi^\dagger (T/2)J_\pi(-T/2)\right>_N
\right|_{N = \bar N}.
\label{fluct_log}
\end{align}
\end{widetext}
Here $\langle ... \rangle_N$ denotes the correlator in an ensemble with a fixed instanton
number $N$ in the volume.
The averaging over the distribution (\ref{dist}) is performed in the narrow-width approximation, by differentiating the fixed-$N$ correlator with respect to $N$; contributions from higher moments of the distribution are 
suppressed by $1/b^2\sim 1/N_c^2$. 

At large Euclidean times the two-point correlator in Eq.~(\ref{fluct_log}) decays exponentially
with the pion mass
\begin{align}
\left<J_\pi^\dagger (T/2)J_\pi(-T/2)\right>_N \sim e^{-m_\pi T}
\end{align}
and the logarithmic derivative becomes
\begin{align}
& \frac{N}{T} \, \frac{d}{dN} \log \left<J_\pi^\dagger (T/2)J_\pi(-T/2)\right>_N 
\nonumber \\
&= - N \frac{dm_\pi}{dN} ,
\end{align}
where $m_\pi$ is understood as a function of the fluctuating instanton number $N$. 
In the instanton vacuum all 
dimensions are fixed by the instanton density $N/V_4=1/R^4$ and the current quark masses. 
The pion mass is the sum of an invariant-like mass and additional explicit chiral symmetry 
breaking and $1/N_c$ contributions,
\begin{align}
m_\pi=\sqrt{m}\bigg(\frac{N}{V_4}\bigg)^{\frac 18}
\left[ C + {\cal O}(mR,1/N_c)\right], 
\end{align}
since it is a Goldstone mode. This implies
\begin{align}
N \frac{d m_\pi}{dN}\; = \frac{m_{0\pi}}{8},
\end{align}
where $m_{0\pi}$ denotes the leading contribution to the pion mass 
in the double limit of small $m$ and large $N_c$.
Altogether, substituting the instanton vacuum result for the correlator 
in Eq.~(\ref{corr}) and cancelling the factors of the normalization volume, 
we obtain
\bea
-\frac{b}{32\pi^2} \, \frac{\langle \pi| F^2 (0) |\pi\rangle}{2 m_\pi} = \frac{m_{0\pi}}{2}.
\label{mass_gluon}
\eea 

The quark contribution to the matrix element (\ref{Trace}) is computed in a similar way from the change of the energy of the pion state under variation of the quark mass (Feynman-Hellmann theorem). The pion sigma term is obtained as
\cite{Zahed:2021fxk}
\begin{align}
\label{FHX}
&\sigma_\pi (0) 
= \frac{\left<\pi|m\overline \psi\psi|\pi\right>}{2m_\pi^2} 
= 
\frac{m}{m_\pi} \frac{dE_\pi}{dm}  = \frac{1}{2},
\end{align}
which agrees with the result obtained from chiral reduction. The quark contribution to the
pion mass is thus
\begin{align}
\frac{\left<\pi|m\overline \psi\psi|\pi\right>}{2m_\pi} = \frac{m_\pi}{2}.
\label{mass_quark}
\end{align}
Together the ILM results 
(\ref{mass_gluon}) and
(\ref{mass_quark})
satisfy the pion mass sum rule (\ref{Trace}) as
\bea
\frac{\left<\pi|T^\mu{}_\mu|\pi\right>}{2m_\pi}
= \left. \frac{m_{0\pi}}{2}\right|_{\rm gluon}
\!\!
+ \left. \frac{m_\pi}{2} \right|_{\rm quark}
\!\!
\approx m_\pi
\eea
in leading order of $m$ and $1/N_c$. One observes that half of the pion mass arises
from the gluonic part of the EMT (conformal anomaly), the other half from the quark part 
(sigma term). This equal splitting is specific to the pion as a Goldstone mode and is 
different in other hadrons such as the nucleon.

An interesting consequence of the mass decomposition (\ref{Trace}) is that the pion matrix 
element of the scalar gluon operator vanishes in the chiral limit,
\begin{align}
\langle \pi | F^2 | \pi \rangle \rightarrow 0 \hspace{1em} (m \rightarrow 0).
\label{F2_chiral_limit}
\end{align}
While this is required by the chiral behavior of the pion mass, it may seem surprising 
from the point of view of hadron structure, since the pion remains a hadron with 
finite size $\sim R$ in the chiral limit (as evidenced e.g.\ by the vector and scalar form factors),
and one would expect the matrix element of $F^2$ to be of the order $\sim R^{-2}$.
The vanishing happens because of the peculiar nature of the Goldstone mode.
In the chiral limit the Goldstone mode decouples from all other (massive) modes of the 
the theory. This includes the $0^{++}$ ``glueball'' modes excited by the the operator $F^2$.
From this point Eq.(\ref{F2_chiral_limit}) expresses the vanishing of the $0^{++}\pi\pi$
coupling in the chiral limit.

It is worth noting that the expectation value of $F^2$ in the pion is negative in~(\ref{mass_gluon}). This is to be understood as the reduction of the 
vacuum gluon condensate in the pion state. When propagating through spacetime, 
the pion thus acts as a ``vacuum cleaner''. In the instanton vacuum
this is realized by the fact that the pion matrix element is proportional 
to the fluctuations of the instanton number, while the average value 
cancels out.

\section{Pion EMT form factors}
\label{PIONEMT}
The general form of the transition matrix element of the EMT between pion states with nonzero
momentum transfer is fixed by Lorentz symmetry, parity, and energy-momentum conservation,
as \cite{Voloshin:1980zf,Donoghue:1991qv}
\begin{align}
\label{THETA123}
&\left<p'\left|T^{\mu\nu}\right|p\right>=\nonumber\\
&2 {\bar{p}^\mu \bar{p}^\nu}\,A(q)+
\frac 12 ({q^\mu q^\nu}-g^{\mu\nu}q^2)\,D(q)
\end{align}
with $q^\mu=(p'-p)^\mu$ and $\bar{p}^\mu=(p+p')^\mu/2$. 
The two form factors correspond to the spin representations
$1\otimes 1=0\oplus 1\oplus 2$, with  $1$ excluded by parity.
It is useful to split the EMT operator according to these 
representations, as the sum of a traceless part 
and a trace part \cite{Ji:1995sv,Ji:2021mtz}, 
\begin{align}
\label{JI}
T^{\mu\nu} &= \left( T^{\mu\nu} - \frac 14 g^{\mu\nu} T^\alpha{}_\alpha \right)
+ \frac 14 g^{\mu\nu} T^\alpha{}_\alpha
\nonumber \\
&\equiv {\bar T}^{\mu\nu} + \frac 14 g^{\mu\nu} T^\alpha{}_\alpha .
\end{align}

At low momenta, the Goldstone nature of
the pion requires~\cite{Voloshin:1980zf,Donoghue:1991qv}
\bea
\label{MOM}
\left<p'\left|T^{\mu\nu}\right|p\right>=p'^\mu p^\nu+p'^\nu p^\mu+\frac 12 g^{\mu\nu} q^2+{\cal O}(p^4),\nonumber\\
\eea
which implies  $D(0)=-1$ \cite{Polyakov:1999gs}. In particular, the form factor of the trace part at low momentum transfer reduces to ($q^2=-Q^2$)
\bea
\label{RADZERO}
\frac{\left<p'\left|T^{\mu}{}_\mu\right|p\right>}{2m_\pi^2}=1-\frac{Q^2}{2m_\pi^2}+{\cal O}(Q^4).
\eea

\subsection{Trace GFF}
The trace part of the pion EMT form factor splits into the
gluonic form factor $G_\pi(q)$ and the pion scalar form factor $\sigma_\pi(q)$,
\bea
\label{MASST}
T_{\pi}(q) \equiv \frac{\langle p'|{T}^\mu{}_\mu|p\rangle}{2m^2_\pi}=G_{\pi}(q)+\sigma_{\pi}(q)
\eea
The scalar form factor is fixed by the strictures of chiral symmetry at low momenta 
and dominated by scalar meson exchange in the instanton vacuum, as detailed in Appendix~\ref{APPSCALAR}. We now address the gluonic form factor.

\subsubsection{Forward}
The forward gluonic matrix element is fixed by the mass identity (\ref{Trace}) and proportional to the vacuum topological compressibility (\ref{COMP}) in the instanton vacuum,
\bea
\label{HVAC}
\frac 1{32\pi^2}\frac{\langle P|F^2|P\rangle}
{2m_\pi^2}\approx -\frac 14 \frac{\sigma_T}{\overline N}
\big(1-\sigma_\pi(0)\big).
\eea
The negative value shows the depletion
of the vacuum gluon condensate in the pion state (see discussion in Sec.~\ref{ANOMALY}).
With this in mind, the forward pion matrix element of the trace of the EMT at leading order in the instanton density
is given by
\begin{widetext}
\bea
\label{MASSX}
T_\pi(0)\approx \big(1-\sigma_\pi(0)\big)\frac b{4\langle F^2\rangle}
\frac 1{32\pi^2}\int d^4x\,\langle F^2(x)F^2(0)\rangle_{\rm conn} +\sigma_\pi(0)
\eea
\end{widetext}
where the correlator is the connected vacuum correlator of the scalar gluon densities at zero momentum transfer.

\begin{figure}
    \centering
    \includegraphics[scale=0.32]{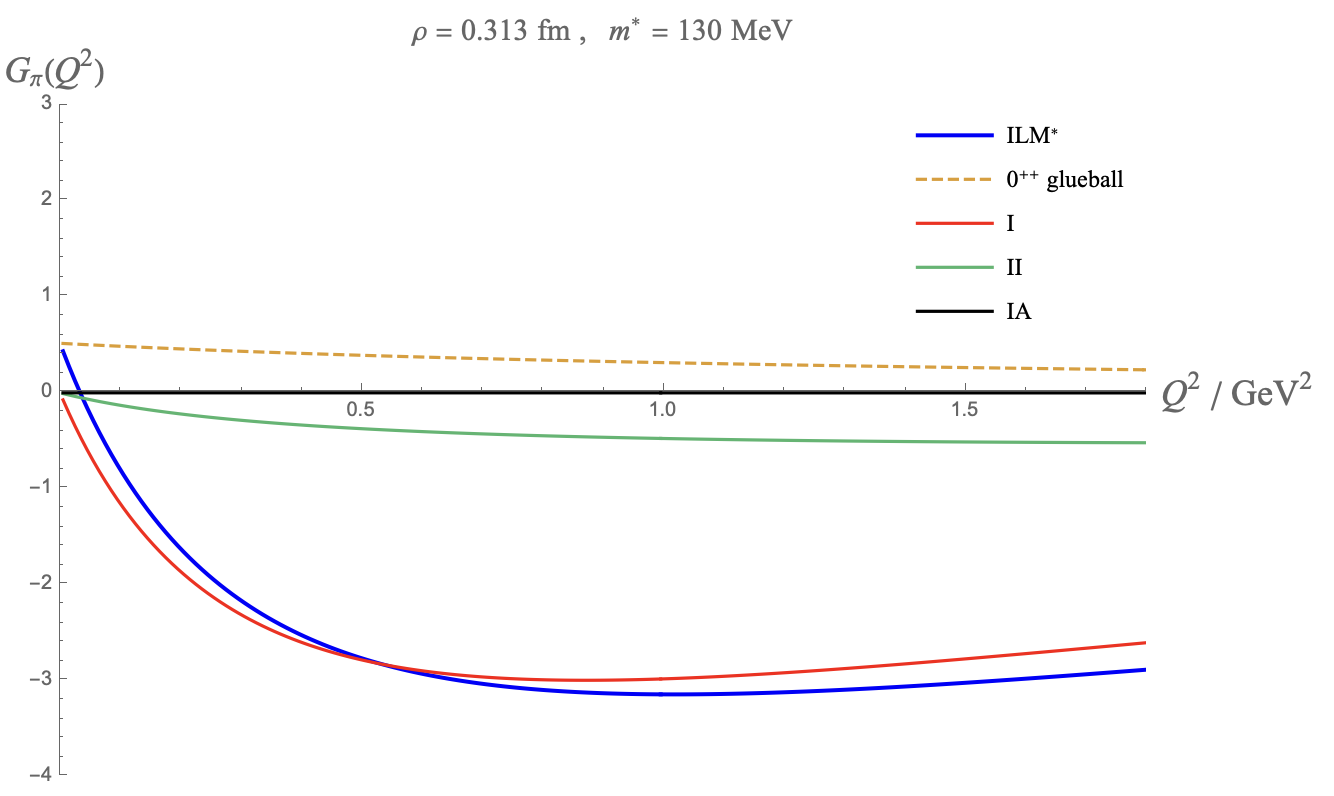}
    \caption{Separate contributions to the gluonic GFF $G_\pi(Q^2)$ in (\ref{MASSZ}) from the ILM at a resolution of $\mu=1/\rho$. }
    \label{fig:GG_contribution}
\end{figure}

\begin{figure*}
%[ht!]
\centering
\subfloat[\label{fig:GG_form_factor}]
{%
\includegraphics[height=5.1cm,width=.48\linewidth]{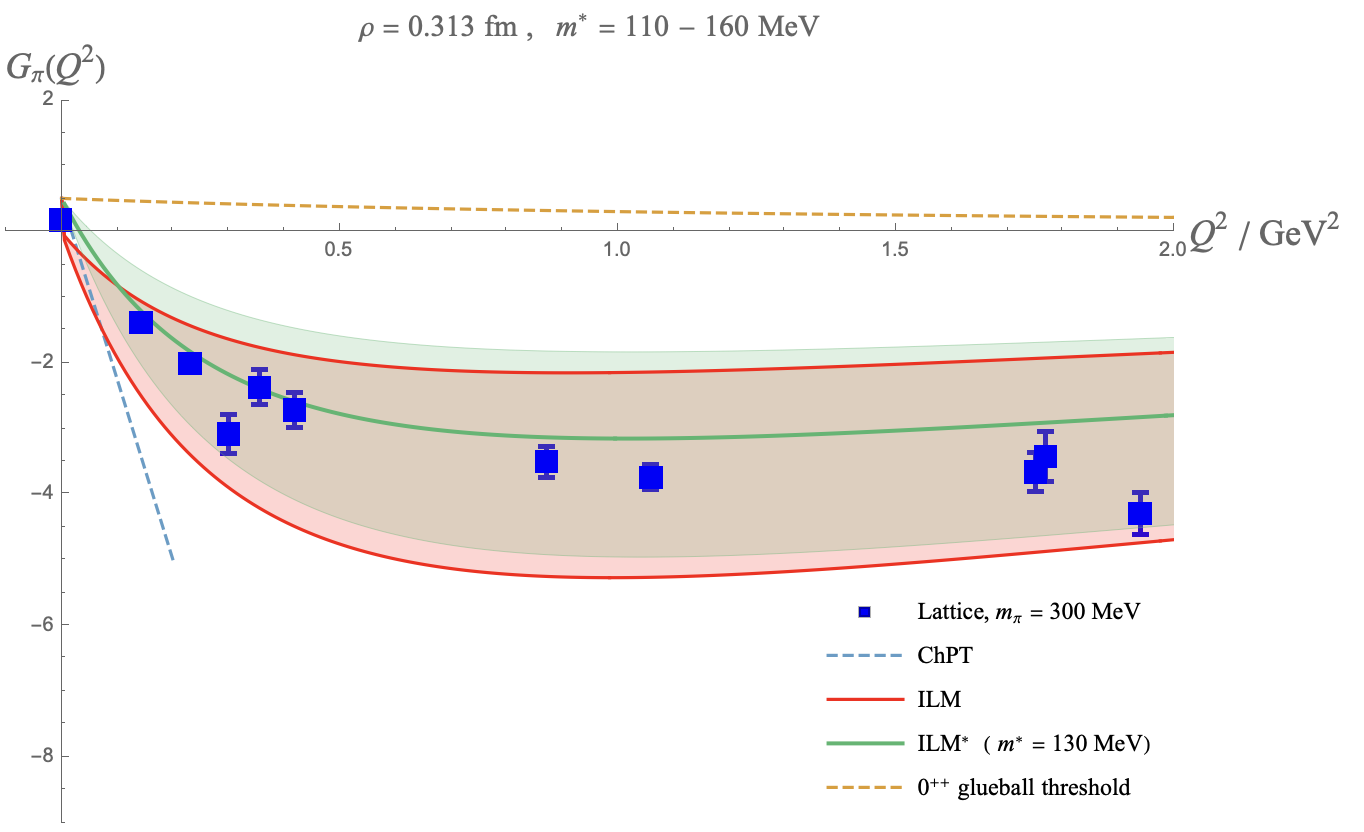}%
}\hfill
\subfloat[\label{fig:Trace_form_factor}]
{%
\includegraphics[height=5.1cm,width=.48\linewidth]{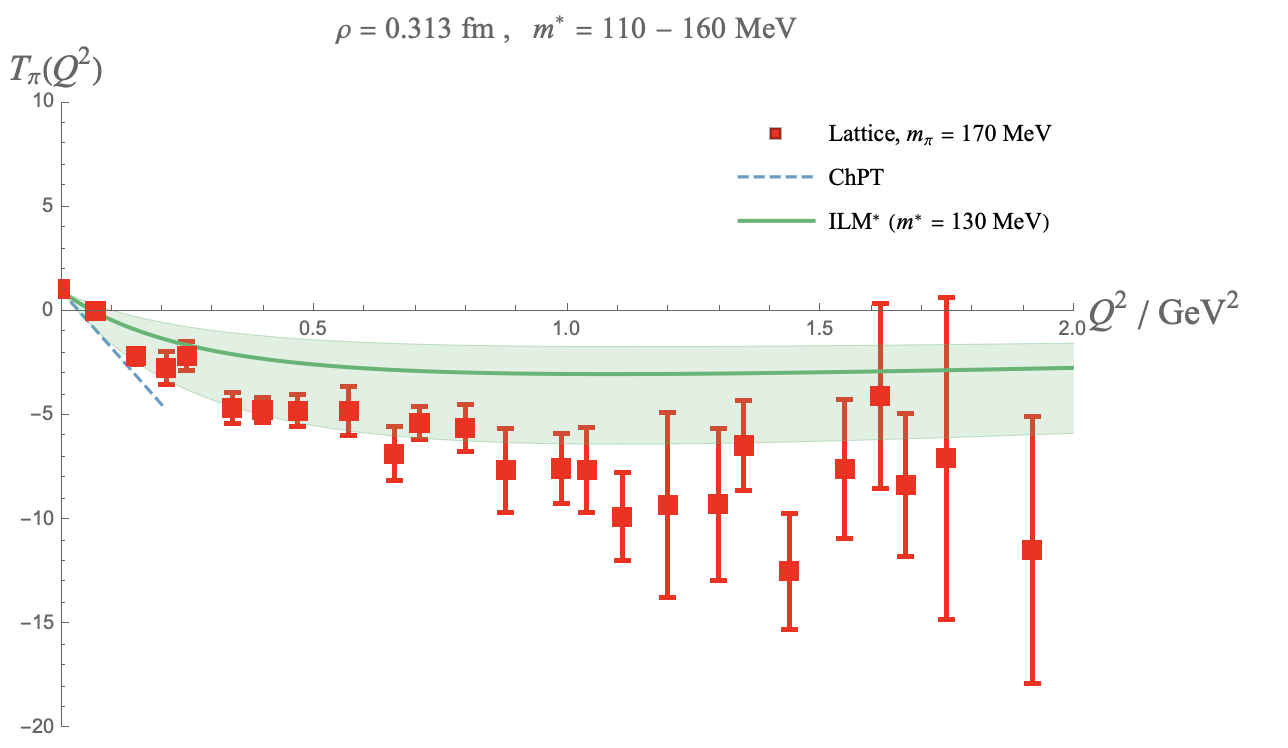}%
}
\caption{Pion gluon scalar form factors. Blue square represents the lattice from \cite{Wang:2024lrm} with pion mass $m_\pi=300$ MeV, the sky blue dashed curve represents the ChPT prediction in Eq.\eqref{RADZERO}. The red bands represents the semi-hard zero mode contribution with the determinantal mass $m^*$ from $110$ to $160$ MeV. The green curve represents the ILM calculation with the bands range the determinantal mass $m^*$ from $110$ to $160$ MeV. The yellow dashed curve represents the $0^{++}$ glueball exchange in the scalar glueball correlation function. See text.}
\label{fig:pion_trace_FF}
\end{figure*}

\subsubsection{Off-forward}
In the gluonic form factor at nonzero momentum transfer one distinguishes between the 
soft regime $Q\rho\ll 1$, where the instantons act collectively in the form of scalar or glueball exchanges, and the semi-hard regime $Q\rho\sim 1$, where single instantons and pairs are resolved~\cite{Shuryak:2020ktq}. The hard regime $Q\rho\gg 1$ of the pion form factor is fixed by factorization and the hard scattering rules.
\\
\\
{\bf Soft regime and transition to semi-hard regime:}
In the soft regime the scalar gluon factor arises from the exchange of scalar mesons
and glueballs with masses $\sim 1/\rho \gg Q$. In this context the $F^2 F^2$ 
vacuum correlator in the forward matrix element (\ref{MASSX})
may be regarded as the ultralocal limit of the momentum-dependent point-to-point 
$F^2 F^2$ correlator \cite{Kacir:1996qn},
\begin{widetext}
\bea
\overline{G}_\pi(0)=&&
\big(1-\sigma_\pi(0)\big)\bigg(\frac b4
\frac{\sigma_T}{\bar{N}}\frac{(2\pi)^4}{V}\delta^4(q)\bigg)\nonumber\\
=&&\big(1-\sigma_\pi(0)\big)\bigg(\frac b4\,
\frac 1{32\pi^2\langle F^2\rangle}\int d^4x\,e^{-iq\cdot x}\,\langle F^2(x)F^2(0)\rangle_{\rm conn} \bigg)_{q=0} .
\eea
\end{widetext}
The momentum-dependent $F^2F^2$ vacuum correlator was analyzed in the ILM using bosonization 
in the soft regime \cite{Kacir:1996qn}, and numerically in the semi-hard regime \cite{Schafer:1996wv}.
It is natural to suppose that the $Q^2$-dependence of the pion gluonic form factor in the transition 
from the soft to the semi-hard regime is governed by the $Q^2$ dependence of the $F^2 F^2$
vacuum correlator. The pion gluonic form factor in this regime is thus dominated by a 
scalar glueball $0^{++}$ with mass $m_{0^{++}}=1.25\,{\rm GeV}$~\cite{Schafer:1996wv}, 
with a small admixture of sigma meson. With this in mind, we have ($q^2=-Q^2$)
\begin{widetext}
\bea
\label{MASSZZ}
\overline{G}_\pi(q)\approx \big(1-\sigma_\pi(0)\big)
\bigg(\frac b{4\langle F^2\rangle}
\frac{\sigma_T}{V_4}\bigg)\bigg(\frac {1}{1+Q^2/{m_{0^{++}}^2}}\bigg) .
\eea
%\end{widetext} 
\\
\\
{\bf Semi-hard regime:}
In this regime the gluonic EMT form factor receives contributions from single instantons and pairs at leading (LO) and next-to-leading (NLO) order in the instanton density, respectively. A detailed calculation in the ILM gives~\cite{Liu:2024}  
%\begin{widetext}
\bea
\label{MASSZ}
G_\pi(q)\approx &&\overline{G}_\pi(q)
+\frac{b}{4}\left[\frac{4}{N_c}\left(\frac{2\kappa}{(\rho m^*)^2}\right)\beta^{(I)}_{2g}(\rho q)+\frac{4}{2N_c(N_c^2-1)}\left(\frac{2\kappa}{(\rho m^*)^2}\right)^2\frac{3m}{m^*}\beta^{(II)}_{2g}(\rho q)\right](\sigma_\pi(q)-\sigma_\pi(0))\nonumber \\
&&+\frac{b}{2N_c(N_c^2-1)}\left(\frac{2\kappa}{(\rho m^*)^2}\right)^2m^2\frac{\rho^4}{9}T_{IA}(\rho m^*)\beta^{(IA)}_{2g}(\rho q)q_\mu q_\nu\frac{\langle p'|\bar{\psi}\left(\gamma_{(\mu}i\overleftrightarrow{\partial}_{\nu)}-\frac{1}{4}g_{\mu\nu}i\overleftrightarrow{\slashed{\partial}}\right)\psi|p\rangle}{2m^2_\pi}\nonumber\\
\eea
again with space-like $q^2=-Q^2$, where the non-local form factors induced by the finite instanton size are defined as
\bea
\label{BETAGG}
    \beta^{(I)}_{2g}(q)&=&\frac{1}{q}\int_0^{\infty} dx\frac{24x^2}{(x^2+1)^4}J_1(qx)
= {q^2 \over 2}K_2(q) \nonumber\\
    \beta^{(II)}_{2g}(q)&=&\frac{1}{q}\int_0^\infty dx\frac{8(2-x^2)}{(1+x^2)^4}J_1(qx)
    = 16/q^2 - 1/2 q^2\big(K_2(q) + 4 K_3( q)\big) \nonumber\\
    \beta^{(IA)}_{2g}(q)&=&\frac{1}{q}\int_0^\infty dx\frac{576 x^2}{(1+x^2)^4}\frac{J_3(qx)}{q^2x^2}
  =   12 \big[48 (-32 + q^2) +  \nonumber \\
 & &  q^2 (768 + 72 q^2 + q^4) K_0( q)
   + 
   12 q (128 + 28 q^2 + q^4)K_1(q)\big]/q^6
\eea
\end{widetext}
and are normalized to unity in the forward limit. The dominant LO instanton contribution
in (\ref{MASSZ}) is driven by the pion scalar form factor, which is detailed in Appendix~\ref{APPSCALAR}.

In Fig.~\ref{fig:GG_contribution} we show the different contributions in (\ref{MASSZ}) to the pion gluonic FF from the ILM. The full result (blue-solid line) is comparable 
to the LO instanton contribution (red-solid line), with the pair II-contribution
(green-solid line) and the $0^{++}$ glueball contribution (orange-dashed line) 
averaging out. The contribution of the IA pairs (black-solid line) is highly suppressed at NLO and about null.

In Fig.~\ref{fig:pion_trace_FF} (left)  we show our result for the behavior of the gluonic scalar form factor $G_\pi(q)$ in the ILM  for a range of determinantal masses
$m^*=110-160\,\rm MeV$ without $\overline{G}_\pi(q)$ (red-spread) and with  $\overline{G}_\pi(q)$ (green-spread). The lattice results (blue squares)~\cite{wang2024trace} are shown along with the result from chiral perturbation theory ChPT (blue-dashed line). The results in the ILM are sensitive to the determinantal mass $m^*=110-160\,\rm MeV$ shown by the band width. Overall, the gluonic part of
the pion EMT is dominated by the instanton LO contribution in the semi-hard regime, with a small contribution from the $0^{++}$ exchange in the soft regime. This is to be compared to our recent result for the gluon
contribution to the nucleon, where the reverse is observed~\cite{Liu:2024}. The reason is the effective suppression of the
glue by the pion mass in the forward direction, due to the Goldstone nature of the pions.

In Fig.~\ref{fig:pion_trace_FF} (right) we show our result for the trace
of the EMT $T_\pi(q)$ in the ILM with the same range of determinantal masses  $m^*=110-150\,\rm MeV$ (green-solid line), versus the reconstructed lattice results (red squares) from
\cite{Hackett:2023nkr}, and the result of ChPT (blue-dashed line). In the reconstruction of the lattice results for $T_\pi(q)$, we made use of the pion gravitational $A,D$ form factors from the lattice in~\cite{Hackett:2023nkr}  and the identity  (the errors were added in quadratures).
 \bea
\label{TT}
%&&\left<p_2\left| T^\mu{}_\mu(0)\right|p_1\right>=2m_\pi^2\nonumber\\
T_\pi(Q^2)=\left(1+\frac{Q^2}{4m_\pi^2}\right) A(Q^2)+\frac{3Q^2}{4m_\pi^2}\, D(Q^2)\nonumber\\
\eea

\subsection{Traceless GFF: gluons}
In LO in the instanton density,
the  traceless part of the gluonic EMT in (\ref{JI}) vanishes,  owing to the self-duality
of the gauge fields. At NLO in the density, it is mostly fixed by the quark energy momentum tensor~\cite{Liu:2024} 
\begin{widetext}
\begin{equation}
\begin{aligned}
\label{TMUNUX}
    \langle p'|g^2\bar{T}^g_{\mu\nu}|p\rangle=&\frac{1}{2N_c(N_c^2-1)}\left(\frac{2\kappa}{(\rho m^*)^2}\right)^2\rho^2m^2T_{IA}(\rho m^*)\\
    &\times\Bigg\{\frac{16\pi^2}{3}\beta^{(IA)}_{T_g,1}(\rho q)\langle p'|\bar{\psi}\left(\gamma_{(\mu} i\overleftrightarrow{\partial}_{\nu)}-\frac{1}{4}g_{\mu\nu}i\overleftrightarrow{\slashed{\partial}}\right)\psi|p\rangle\\
    &-\frac{4\pi^2\rho^2}{9}\beta^{(IA)}_{T_g,2}(\rho q)\left(q_\mu q_\rho g_{\nu\lambda}+q_\nu q_\rho g_{\mu\lambda}-\frac{1}{2} g_{\mu\nu}q_\rho q_\lambda\right)\langle p'|\bar{\psi}\left(\gamma_{(\rho}i\overleftrightarrow{\partial}_{\lambda)}-\frac{1}{4}g_{\rho\lambda}i\overleftrightarrow{\slashed{\partial}}\right)\psi|p\rangle\\
    &-4\pi^2\rho^4\beta^{(IA)}_{T_g,3}(\rho q)\left(q_\mu q_\nu-\frac{1}{4}q^2g_{\mu\nu}\right)q_\rho q_\lambda\langle p'|\bar{\psi}\left(\gamma_{(\rho}i\overleftrightarrow{\partial}_{\lambda)}-\frac{1}{4}g_{\rho\lambda}i\overleftrightarrow{\slashed{\partial}}\right)\psi|p\rangle\Bigg\}\\
    &+\frac{1}{2N_c(N_c^2-1)}\left(\frac{2\kappa}{(\rho m^*)^2}\right)^2\frac{m}{m^*}\frac{8\pi^2\rho^2}{9}\beta^{(II)}_{T_g}(\rho q)\left(q_\mu q_\nu-\frac{1}{4}q^2g_{\mu\nu}\right)\langle p'|m\bar{\psi}\psi|p\rangle
\end{aligned}
\end{equation}
where the non-local form factors are defined as
\begin{equation}
\label{BETAEMT}
    \beta^{(IA)}_{T_g,1}(q)=\frac{1}{q}\int_0^\infty dx\left[\frac{24}{(1+x^2)^4}J_1(qx)+\frac{24x^2}{(1+x^2)^4}J_3(qx)-\frac{192}{(1+x^2)^3}\frac{J_3(qx)}{q^2x^2}\right]
\end{equation}
\begin{equation}
    \beta^{(IA)}_{T_g,2}(q)=\frac{1}{q}\int_0^\infty dx9x^2\left[\frac{128x^2}{(1+x^2)^4}\frac{J_3(qx)}{q^2x^2}-\frac{512}{(1+x^2)^3}\frac{J_4(qx)}{q^3x^3}\right]
\end{equation}
\begin{equation}
    \beta^{(IA)}_{T_g,3}(q)=\frac{1}{q}\int_0^\infty dx\frac{256x^4}{(1+x^2)^3}\frac{J_5(qx)}{q^4x^4}
\end{equation}
\begin{equation}
    \beta^{(II)}_{T_g}(q)=\frac{1}{q}\int_0^\infty dx\frac{576x^2}{(1+x^2)^4}\frac{J_3(qx)}{q^2x^2}
\end{equation}
and are normalized to 1
at $q = 0$. The quark hopping integral between the instanton and (anti)-instanton $T_{IA}(\rho m^*)$ is defined as
\bea
\label{QHOPPING}
T_{IA}(\rho m^*)=\frac{1}{16\pi^2}\int d^4R\int_0^\infty dk\frac{k^3\mathcal{F}(k)J_2(kR)}{k^2+(\rho m^*)^2}
\eea
Using (\ref{THETA123}) for the invariant GFFs of the quark and gluon EMT,
\bea
\label{THETA123QG}
\left<p'\left|\bar{T}_{q,g}^{\mu\nu}\right|p\right>&=&
2 {\bar{p}^\mu \bar{p}^\nu}\,A_{q,g}(q)+
\frac 12 ({q^\mu q^\nu}-g^{\mu\nu}q^2)\,D_{q,g}(q)\nonumber\\
&&- \frac{1}{4} g^{\mu\nu}\left[\left(1-\frac{q^2}{4m_\pi^2}\right) A_{q,g}(q)-\frac{3q^2}{4m_\pi^2}\, D_{q,g}(q)\right],
\eea
we can express the gluon GFFs in terms of the quark GFFs with the help of (\ref{TMUNUX}), with the result
\begin{equation}
\label{AGQ}
A_g(Q^2)=\frac{1}{2N_c(N_c^2-1)}\left(\frac{2\kappa}{(\rho m^*)^2}\right)^2\rho^2m^2T_{IA}(\rho m^*)\frac{16\pi^2}{3}\beta^{(IA)}_{T_g,1}(\rho q)A_q(Q^2) 
\end{equation}
\begin{equation}
\begin{aligned}
\label{DGQ}
D_g(Q^2)=&\frac{1}{2N_c(N_c^2-1)}\left(\frac{2\kappa}{(\rho m^*)^2}\right)^2\rho^2m^2T_{IA}(\rho m^*)\\
\times\bigg\{&\left[\frac{16\pi^2}{3}\beta^{(IA)}_{T_g,1}(\rho q)+\frac{2\pi^2}{3}\rho^2Q^2\beta^{(IA)}_{T_g,2}(\rho q)-3\pi^2\rho^4Q^4\beta^{(IA)}_{T_g,3}(\rho q)\right]D_q(Q^2)\\
&+\frac{8\pi^2}{9}\rho^2m_\pi^2\left[\beta^{(IA)}_{T_g,2}(\rho q)-\frac{9}{2}\rho^2Q^2\beta^{(IA)}_{T_g,3}(\rho q)\right]\left(1+\frac{Q^2}{4m_\pi^2}\right)A_q(Q^2)\bigg\}\\
&+\frac{1}{2N_c(N_c^2-1)}\left(\frac{2\kappa}{(\rho m^*)^2}\right)^2\frac{m}{m^*}\frac{32\pi^2}{9}\rho^2m_\pi^2\beta^{(II)}_{T_g}(\rho q)\sigma_\pi(Q^2)
\end{aligned}
\end{equation}
\end{widetext}

\begin{figure*}
\centering
\subfloat[\label{fig:A_qg}]
{%
\includegraphics[height=6cm,width=.47\linewidth]{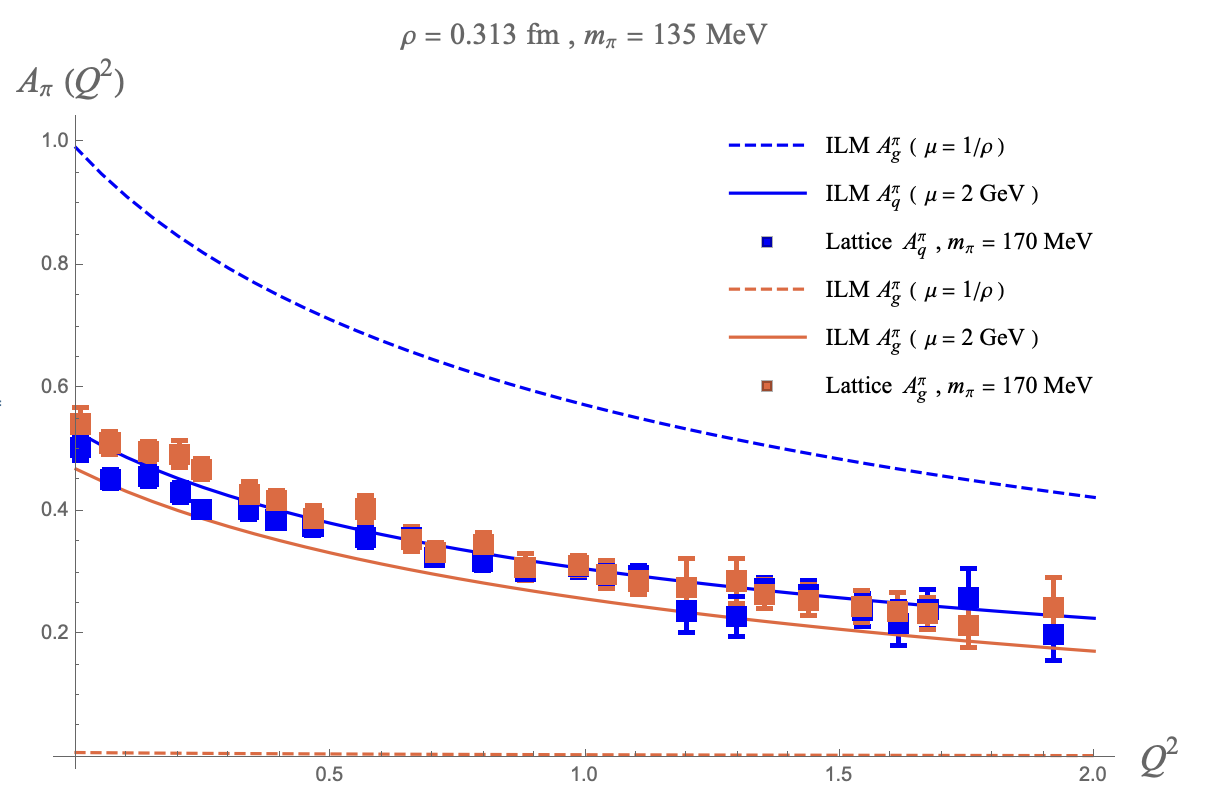}}%
\hfill
\subfloat[\label{fig:D_qg}]
{%
\includegraphics[height=6cm,width=.47\linewidth]{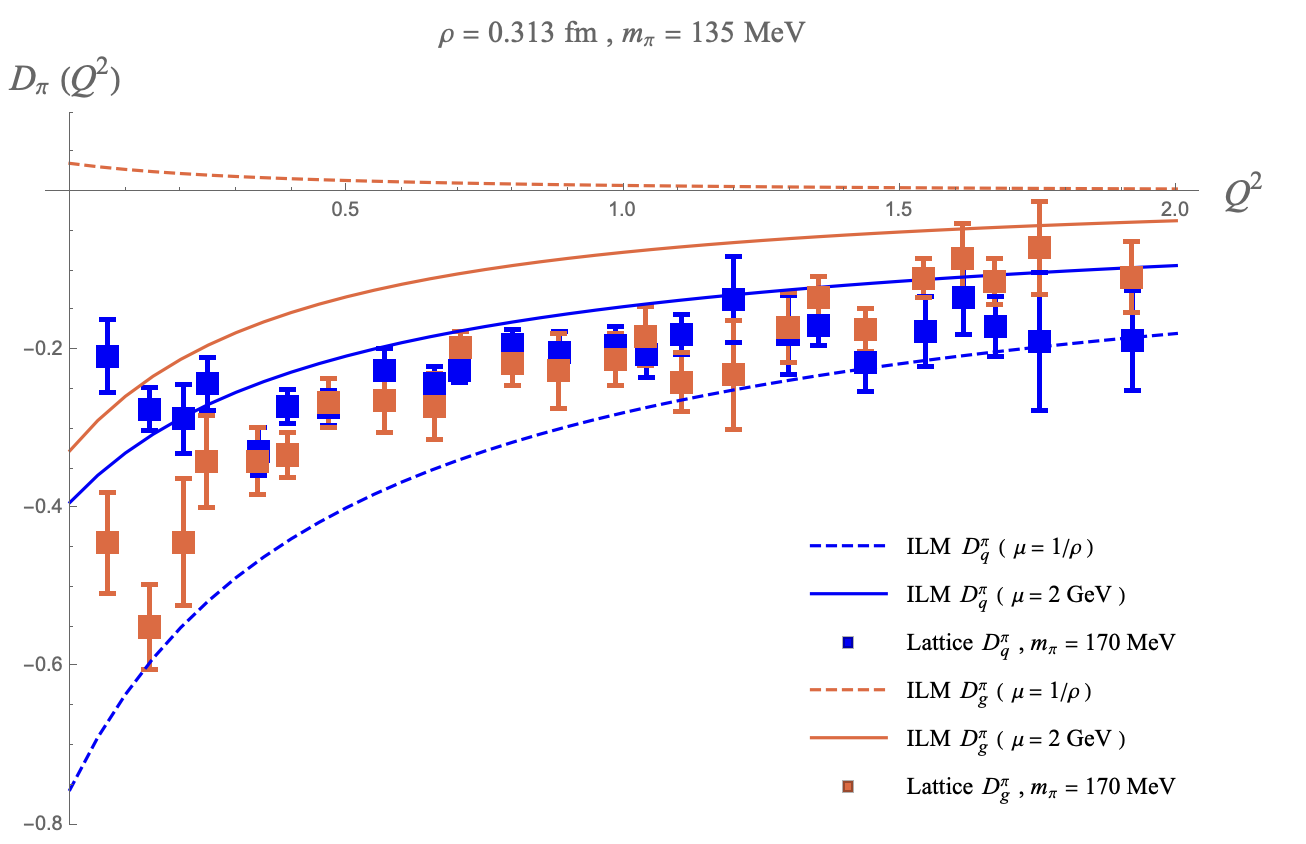}}%
 \caption{a: Quark contribution $A_q(Q)$ (blue) and gluon contribution $A_g(Q)$ (orange) to the pion GFF in the ILM at a resolution $\mu=1/\rho$ (dashed lines) and at a resolution $\mu=2\,\rm GeV$ (solid lines). The lattice results are from~\cite{Hackett:2023nkr}. 
  b: Quark contribution $D_q(Q)$ (blue) and gluon contribution $D_g(Q)$ (orange) to the pion GFF in the ILM at a resolution $\mu=1/\rho$ (dashed lines) and at a resolution $\mu=2\,\rm GeV$ (solid lines). The lattice results are from~\cite{Hackett:2023nkr}.}
\end{figure*}

The gluonic contribution to the pion gravitational charge $A_g(0)$ and $D_g(0)$ receives a contribution from the instanton-antiinstanton molecules at NLO
in (\ref{AGQ}-\ref{DGQ}). Using the QCD instanton vacuum parameters
\cite{Liu:2023fpj,Liu:2023yuj} (and references therein), we have
at the resolution $\mu=1/\rho$
\bea
A_q(0)&=&0.9924\qquad\,\,\,\,\, A_g(0)=0.0076\nonumber\\
D_q(0)&=&-0.7539\qquad D_g(0)=0.03556\nonumber\\
\eea
At the resolution scale $\mu=1/\rho$, the emergent quark contributions to the pion
GFFs are much larger than the gluon contributions, as the latter are penalized by powers
of the instanton density. These estimates are consistent with those given in~\cite{Zahed:2021fxk}. However, it is important to note that the emergent 
constituent quarks arise solely from the glue carried by the solitonic
instantons in the ILM!
%Note that $\bar{C}_q(t)$ and $\bar{C}_g(t)$ do not appear.

To compare with the lattice, we can evolve the result from $\mu_0=0.6~\mathrm{GeV}\sim1/\rho$, the normalization scale of the instanton vacuum \cite{Liu:2023fpj,Liu:2023yuj}, using the DGLAP equation~\cite{Hatta:2018sqd}
\bea
\label{DGLAP}
   && \mu\frac{d}{d\mu}\begin{pmatrix}
    A_q(0) \\
    A_g(0)
    \end{pmatrix}_\mu=\nonumber\\
    &&\frac{\alpha_s}{4\pi}\begin{pmatrix}
    -\frac{16}{3}C_F & \frac{4}{3}N_f\\
    \frac{16}{3}C_F & -\frac{4}{3}N_f
    \end{pmatrix}
    \begin{pmatrix}
    A_q(0) \\
    A_g(0)
    \end{pmatrix}_\mu ,
\eea
where $C_F=\frac{N_c^2-1}{2N_c}$. The evolution is the same for $D_q(0)$ and $D_g(0)$.
%At $\mu=2$ GeV, the results are shown in the table

In Fig.~\ref{fig:A_qg} we show the separate quark and gluon contributions to the pion GFF $A$ in the ILM. At 
the resolution $\mu=1/\rho$, the quark contribution $A_q(Q^2)$ (blue-dashed line) dwarfs the gluon contribution $A_g(Q^2)$ (orange-dashed line). Under 
QCD evolution of the $Q=0$ charges to the resolution
$\mu=2\,\rm GeV$, the quark contribution $A_q(Q^2)$ 
(blue solid line) becomes comparable to the 
gluon contribution $A_g(Q^2)$ (orange solid line). They
are in good agreement with the 
lattice results for $A_q(Q^2)$ (blue squares) and 
$A_g(Q^2)$ (orange squares) at the scale $\mu=2\,\rm GeV$~\cite{Hackett:2023nkr}. The same
results and comparison for the quark and gluon contributions to the pion D-GFF are shown in Fig.~\ref{fig:D_qg}.

\begin{widetext}

\subsection{Traceless GFF: quarks}
The  quark contribution to the traceless part of the EMT ${\bar T}^{\mu\nu}_q$
in the instanton vacuum is dominant in LO in the instanton density. On the light front, it can be estimated  
using the in-out states given by (\ref{PWF}). 
In the Drell-Yan frames with $q^+=0$,  a collection of frames related by light-front boosts, we specialize to the Breit frame with $p_\perp=0$, where the pion momenta have light-front components
$$p_1^\mu=\bigg(P^+, -\frac 12 q_\perp, \frac{m_\pi^2+\frac 14 q_\perp^2}{2P^+}\bigg)\qquad\qquad
p_2^\mu=\bigg(P^+, +\frac 12 q_\perp, \frac{m_\pi^2+\frac 14 q_\perp^2}{2P^+}\bigg)
$$
The ensuing form factors are real and free of Z-contributions. In particular, we have
\iffalse
\bea
\label{T+-LFX}
&&{\langle p_2|{\bar T}_q^{-+}(0)|p_1\rangle}
%{2m_\pi^2}
=\int dx\int \frac{d^2k_\perp}{(2\pi)^3}\nonumber\\
&&\times\bigg(
\Phi^*_\pi(x, k_\perp+\bar{x}q_\perp, s_1', s_2')\,
{\overline{u}_{s_1'}(k+q)\gamma^+k^-u_{s_1}(k)}\,\frac{\delta_{s_2's_2}}{4x}\,
\Phi_\pi(x, k_\perp, s_1, s_2)\nonumber\\
&&\qquad+\Phi^*_\pi(x, k_\perp-{x}q_\perp, s_1', s_2')\,
{\overline{v}_{s_2}(k)\gamma^+k^-v_{s_2'}(k+q)}\,
\frac{\delta_{s_1's_1}}{4\bar x}\,
\Phi_\pi(x, k_\perp, s_1, s_2)\bigg)
\eea
%\end{widetext}
and
\fi
\bea
\label{T++LFX}
&&\frac{\langle p_2|{\bar T}_q^{++}(0)|p_1\rangle}
{2P^{+2}}=\frac 1{2P^+}\int dx\int \frac{d^2k_\perp}{(2\pi)^3}\nonumber\\
&&\times\bigg(
\Phi^*_\pi(x, k_\perp+\bar{x}q_\perp, s_1', s_2')\,
{\overline{u}_{s_1'}(k+q)\gamma^+u_{s_1}(k)}\,\frac{\delta_{s_2's_2}}{2}\,
\Phi_\pi(x, k_\perp, s_1, s_2)\nonumber\\
&&\qquad+\Phi^*_\pi(x, k_\perp-{x}q_\perp, s_1', s_2')\,
{\overline{v}_{s_2}(k)\gamma^+v_{s_2'}(k+q)}\,
\frac{\delta_{s_1's_1}}{2}\,
\Phi_\pi(x, k_\perp, s_1, s_2)
\bigg)
\eea

More specifically,  we have
\iffalse
\bea
\label{T+-LFXX}
&&{\langle p_2|{\bar T}_q^{+-}(0)|p_1\rangle}
%{2m_\pi^2}
= \int dx\int \frac{d^2k_\perp}{(2\pi)^3}\nonumber\\
&&\times\bigg( \phi_\pi(x, k_\perp+\bar x q_\perp)\phi_\pi(x, k_\perp)
\bigg(\frac{k_\perp^2+M^2+\bar x k_\perp\cdot q_\perp}{ x}\bigg)\nonumber\\
&&\qquad+\phi_\pi(x, k_\perp- x q_\perp)\phi_\pi(x, k_\perp)
\bigg(\frac{k_\perp^2+M^2- x k_\perp\cdot q_\perp}{\bar x}\bigg)
\bigg)\bigg(\frac{k_\perp^2+M^2}{x\bar x}\bigg)
\eea
and
\fi

\bea
\label{T++LFXX}
&&\frac{\langle p_2|{\bar T}_q^{++}(0)|p_1\rangle}
{2P^{+2}}= \int dx\int \frac{d^2k_\perp}{(2\pi)^3}\nonumber\\
&&\times\bigg( \phi_\pi(x, k_\perp+\bar x q_\perp)\phi_\pi(x, k_\perp)
4x\bigg(\frac{k_\perp^2+M^2+\bar x k_\perp\cdot q_\perp}{x\bar x}\bigg)\nonumber\\
&&\qquad+\phi_\pi(x, k_\perp- x q_\perp)\phi_\pi(x, k_\perp)
4\bar{x}\bigg(\frac{k_\perp^2+M^2-x k_\perp\cdot q_\perp}{x\bar x}\bigg)
\bigg)
\eea

\begin{figure*}
%[ht!]
\centering
{%
\includegraphics[height=5cm,width=.47\linewidth]{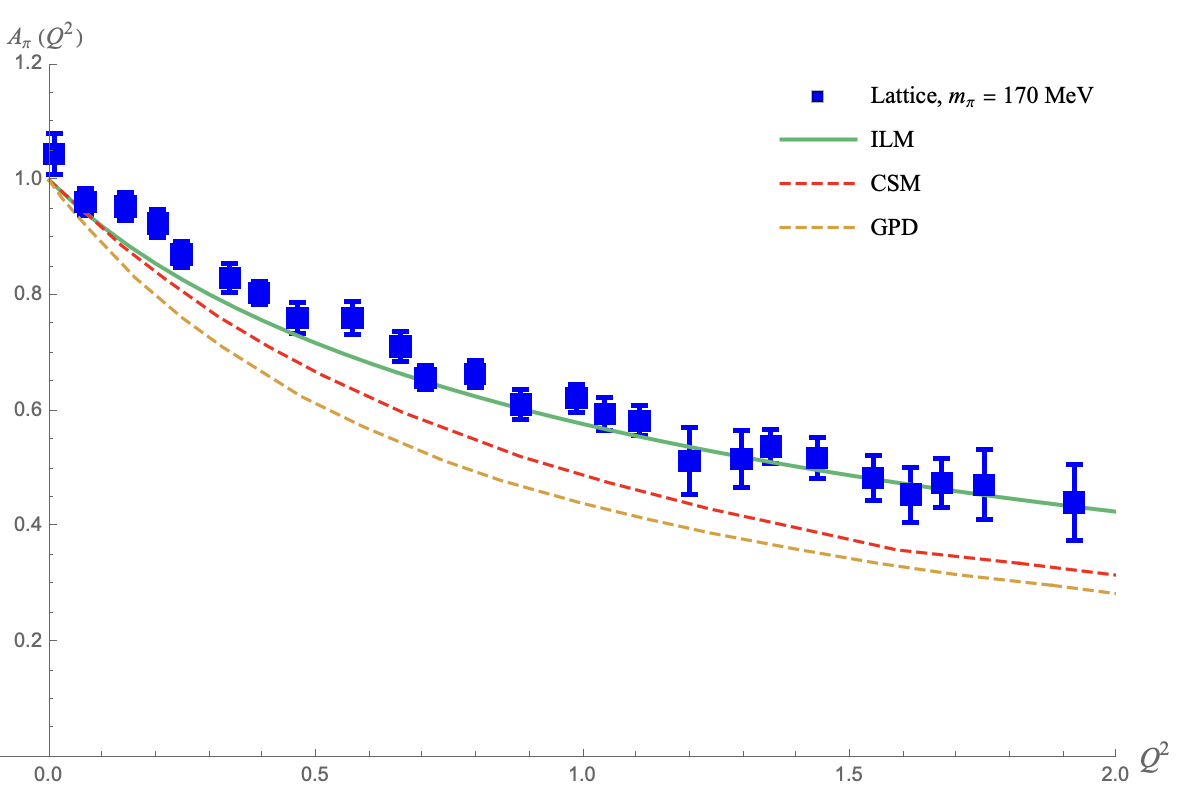}%
}\hfill
{%
\includegraphics[height=5cm,width=.47\linewidth]{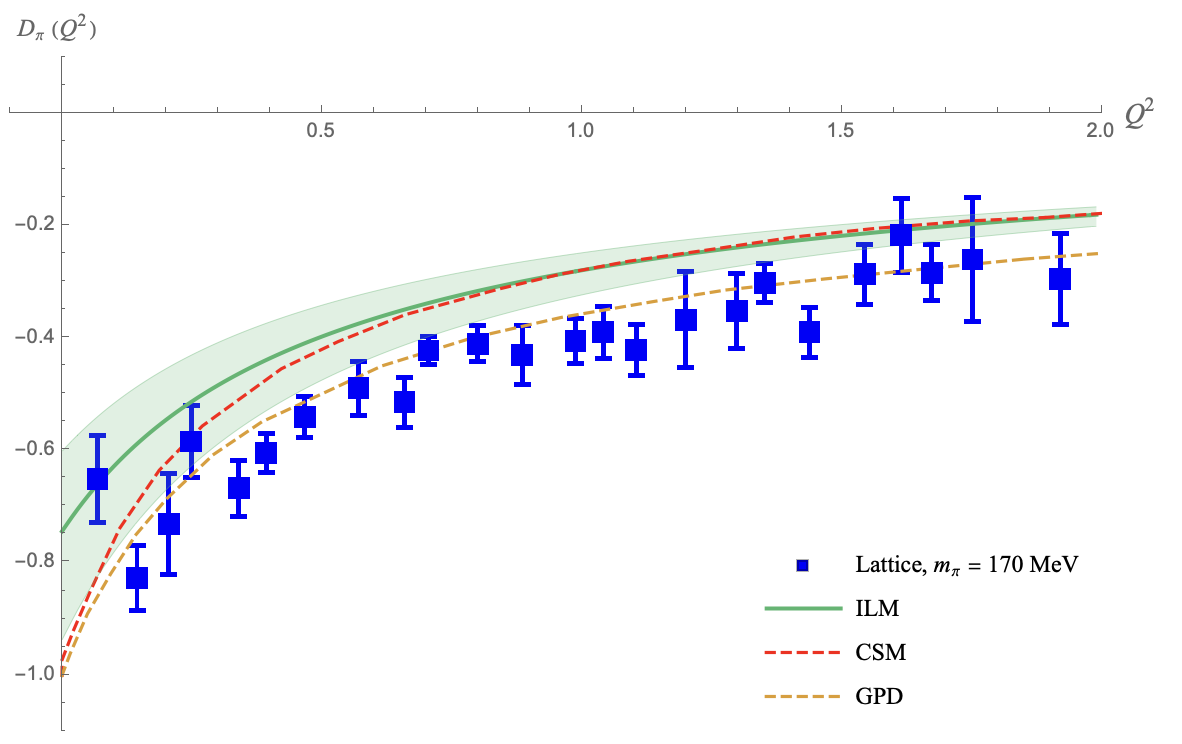}%
}
\caption{Pion form factors $A(Q^2)$ (left green line) and $D(Q^2)$
(right green line) in the ILM, compared to the lattice results~\cite{Hackett:2023nkr}. GPD denotes the method using algebraic GPD ansatz in \cite{Raya:2021zrz} and Bethe-Salpeter based method with continuum Schwinger functions (CSM) \cite{Xu:2023izo}}
\label{fig:AD_PIONFF}
\end{figure*}

We recall that on the light front Lorentz symmetry is 
only recovered when the multi-Fock space is fully retained. 
It is approximate when it is truncated, although the 
underlying dynamics may turn it to a fair approximation,
e.g. see the recent analysis for the longitudinal and transverse rho meson in the QCD instanton vacuum~\cite{Liu:2023fpj}. In  general, 
the Lorentz structures (\ref{THETA123}), when applied to the 2-particle Fock component (\ref{PWF}), may require further amendments. Indeed, additional and spurious form factors do emerge, that depend
explicitly on the light front null vector $n^\mu=(1,0_\perp, -1)$~\cite{Carbonell:1998rj}. Fortunately, the spurious  form factors drop out from the manifestly physical components $T^{++}$ (the light front momentum) and  $T^{-+}$ (the light front Hamiltonian).

\section{Pion GFFs}
\label{RESULTS}
The pion GFFs $A$ and $D$ in (\ref{THETA123}) capture the mass distributions inside the pion. They are obtained by solving the expressions for the light-front components
%\begin{widetext}
\bea
\label{MOM2X}
\left<p_2|T^{++}|p_1\right>&=&2P^{+2}A(Q^2)\nonumber\\
\left<p_2|T^{-+}|p_1\right>&=&
\bigg(m_\pi^2+\frac 14 Q^2\bigg)A(Q^2)+\frac 12 Q^2D(Q^2)\nonumber\\
\left<p_2|T^\mu{}_\mu|p_1\right>&=&
2 \bigg(m_\pi^2+\frac 14 Q^2\bigg)A(Q^2)+\frac 32 Q^2D(Q^2)
\eea
for the EMT form factors 
\bea
\label{AQDQ}
A(Q^2)&=&\frac 4{Q^2+4m_\pi^2}\bigg(
3\left< p_2|T^{-+}|p_1\right> 
-\left< p_2|T^\mu{}_\mu|p_1\right>
\bigg)\nonumber\\
D(Q^2)&=&
\frac {1}{Q^2}\bigg(\left<p_2|T^\mu{}_\mu|p_1\right>
-2\left<p_2|T^{-+}|p_1\right>\bigg)
\eea
or equivalently, 
\bea
\label{AQDQX}
A(Q^2)&=&\frac 1{2P^{+2}}
\left< p_2|T^{++}|p_1\right> \nonumber\\
D(Q^2)&=&
\frac {2}{3Q^2}\bigg(\left<p_2|T^\mu{}_\mu|p_1\right>
-\frac{Q^2+4m_\pi^2}{4P^{+2}}\left<p_2|T^{++}|p_1\right>\bigg)
\eea
\end{widetext}
 We evaluate $A$ and $D$ from (\ref{AQDQX}), using the fact that $T^{++}$ is a leading twist-2 operator, while $T^{-+}$ is a sub-leading twist-3 operator, asymptotically. The trace of the energy momentum tensor was analyzed earlier in the 
 pion rest frame (see Sec.~\ref{ANOMALY}); as a scalar the results carry over verbatim to the light front frame.

In Fig.~\ref{fig:AD_PIONFF} (left) we show  our result for the pion GFF $A$ in the ILM (green-solid line) in comparison to the lattice results (blue squares)~\cite{Hackett:2023nkr}. We have used the standard ILM parameters given earlier, with a mean instanton size $\rho=0.313\,\rm fm$, a  constituent quark mass
$M=398\,\rm MeV$, and a pion mass $m_\pi=135\,\rm MeV$. The  agreement of the ILM and lattice results is excellent. In Fig.~\ref{fig:AD_PIONFF} (right) we show our result
for the pion $D$ GFF (green band) versus the lattice results (blue squares)~\cite{Hackett:2023nkr}. The ILM results are sensitive to the quark
determinantal mass $m^*=110-160\,\rm MeV$ through the trace part of the EMT,
with most contributions stemming from the semi-hard regime, as noted earlier.

The GFFs $A$ and $D$ are characterized by the radii
\bea
\label{RADII_A}
\langle r^2_{A}\rangle&=&-\frac 6{A(0)}
\bigg(\frac{dA(Q^2)}{dQ^2}\bigg)_{Q=0}\nonumber\\
\langle r^2_{D}\rangle&=&-\frac 6{D(0)}
\bigg(\frac{dD(Q^2)}{dQ^2}\bigg)_{Q=0}
\eea
which are listed in Table~\ref{TAB_1}. They are a measure of the 
range of the tensor and scalar exchange mechanisms in the form factors. Alternatively, 
the scalar and mass radii can be obtained from the pion matrix elements of $T^\mu{}_\mu$ in (\ref{MOM2X}) and $T^{00}$ in the Breit frame, 
\bea
\label{RADII}
\langle r^2_{M}\rangle=-\frac 6{T^\pi_{00}(0)}
\bigg(\frac{dT^\pi_{00}(Q^2)}{dQ^2}\bigg)_{Q=0}\nonumber\\
\langle r^2_{S}\rangle=-\frac 6{T^\pi_{\mu\mu}(0)}
\bigg(\frac{dT^\pi_{\mu\mu}(Q^2)}{dQ^2}\bigg)_{Q=0}
\eea
with 
\bea
\label{energy_and_trace_form_factor}
T^\pi_{\mu\mu}(Q^2)&=&2 \bigg(m_\pi^2+\frac 14 Q^2\bigg)A(Q^2)+\frac 32 Q^2D(Q^2)
\nonumber\\
T^\pi_{00}(Q^2)&=&2 \bigg(m_\pi^2+\frac 14 Q^2\bigg)A(Q^2)+\frac 12 Q^2D(Q^2)
\nonumber\\
\eea
The scalar and mass radii are related to $r_A$ through 
\begin{eqnarray}
    \langle r_M^2\rangle& =&\langle r_A^2\rangle-\frac{3}{2m_\pi^2}\left[1+3D(0)\right] \nonumber\\ 
    \langle r_S^2\rangle& =&\langle r_A^2\rangle -\frac{3}{2m_\pi^2}\left[1+D(0)\right]\nonumber\\ 
\end{eqnarray}
They are listed in~\ref{TAB_2}. The lattice radii from~\cite{Hackett:2023nkr} are extracted from a monopole
fit. The chiral perturbation prediction for the scalar radius follows from \eqref{RADZERO} with $m_\pi=135$ MeV,
\begin{equation}
    \langle r_S^2\rangle_{\mathrm{ChPT}}=\frac{3}{m_\pi^2}
\end{equation}
With the exception of the mass radius, our results are
consistent with the lattice results.

\begin{table}[!h]
\begin{center}
            \caption{Pion A,D gravitational radii.}
    \begin{tabular}{|c|c|c|}
    \hline
     & $\langle r^2_{A}\rangle$ &  $\langle r^2_{D}\rangle$ \\
    \hline
    ILM &  0.216 fm$^2$  & 0.422 fm$^2$    \\
    CSM \cite{Xu:2023izo}&  0.47 fm$^2$ & 0.81 fm$^2$ \\
    GPD \cite{Raya:2021zrz}&  0.56 fm$^2$ & 0.81 fm$^2$ \\
    MD~\cite{Broniowski:2024oyk}&  0.38 fm$^2$ & 0.71 fm$^2$ \\
    Lattice \cite{Hackett:2023nkr}&  0.166 fm$^2$ & 0.376 fm$^2$ \\
    \hline
    \end{tabular}
    \label{TAB_1}
\end{center}
\end{table}

\begin{table}[!h]
 \begin{center}
          \caption{Pion scalar and mass radii.}
    \begin{tabular}{|c|c|c|}
    \hline
     & $\langle r^2_{S}\rangle$ & $\langle r^2_{M}\rangle$ \\
    \hline
    ILM &  4.296 fm$^2$  & -0.560 fm$^2$   \\
    ChPT & 6.407 fm$^2$ & -- \\
    Lattice \cite{Hackett:2023nkr} & 5.385 fm$^2$ & -0.230 fm$^2$ \\
    \hline
    \end{tabular}
    \label{TAB_2}
\end{center}
\end{table}

\begin{figure}
    \centering
    \includegraphics[scale=0.47]{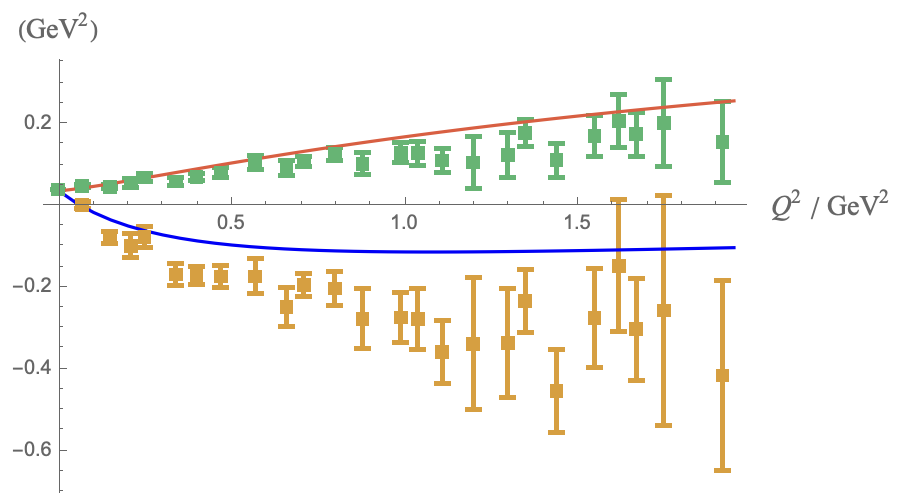}
    \caption{The red curve represents $T^\pi_{00}(Q^2)$ and the blue curve represents$T^\pi_{\mu\mu}(Q^2)$. The green square represents the lattice data for $T^\pi_{00}(Q^2)$ and the yellow square represents the lattice data for $T^\pi_{\mu\mu}(Q^2)$ from \cite{Hackett:2023nkr} }
    \label{fig:radius_FF}
\end{figure}

\section{Conclusion}
\label{CONCLUSION}
In QCD the breaking of conformal symmetry puts stringent constraints on the bulk hadronic correlations in the form of low-energy theorems~\cite{Novikov:1981xi}. In the instanton vacuum these constraints are enforced in the form of stronger-than-Poisson fluctuations in the number of quasiparticles, with a vacuum compressibility 
indicative of a quantum liquid. 

The instanton vacuum states that the gluon part of the trace anomaly contributes 
half the pion mass, with the other half coming from the quark sigma term. 
This constrains the pion form factors of the gluon and quark part of the 
trace of the EMT at zero momentum transfer. The momentum transfer dependence
arises from a short-distance contribution from scalar glueball exchange, 
and a long-distance contribution stemming from the pion scalar form factor, which 
is mostly a 2-pion cloud. The core gluon radius of the pion is about one third 
that of the scalar pion cloud. 

For the traceless part of the EMT, the leading light-front components of the matrix element
have been evaluated using light-front wave functions of the pion derived from the instanton vacuum.
Combined with the results for the trace part of the EMT, this provides a complete characterization 
of the two invariant EMT pion form factors with spin-2 (tensor) and spin-0 (scalar).
The results compare well with  the recent lattice simulations~\cite{Hackett:2023nkr,wang2024trace}.

The gluonic structure of the pion can be probed in exclusive photo/electroproduction of heavy quarkonia on the nucleon with pion knockout, $\gamma N \rightarrow Q\bar Q + \pi + N$, by selecting kinematics where the scattering process takes
place on a peripheral pion in the nucleon (impact parameters $\sim 1/m_\pi$) \cite{Strikman:2003gz}.
The relevance of our results to such measurements is to be explored.

In this work we have studied the transition matrix elements of the EMT between pion states at spacelike momentum transfers. The two-pion creation matrix elements $\langle \pi\pi | F^2 |0\rangle$ and 
$\langle \pi\pi | m\bar\psi\psi  |0\rangle$ at timelike momentum transfers appear in 
the description of hadronic $\tau$ lepton decays $\tau \rightarrow l \pi\pi$ and have been analyzed 
in dispersion theory \cite{Celis:2013xja,Donoghue:1990xh}. The spacelike form factors estimated in the instanton vacuum can constrain
the behavior of the timelike form factors through crossing and analyticity.

\vskip 0.5cm
{\noindent\bf Acknowledgements}

\noindent 
We thank Fangcheng He for a discussion on his lattice results.
This work is supported by the Office of Science, U.S. Department of Energy under Contract  No. DE-FG-88ER40388.
This research is also supported in part within the framework of the Quark-Gluon Tomography (QGT) Topical Collaboration, under contract no. DE-SC0023646.

This material is based upon work supported by the U.S.~Department of Energy, Office of Science,
Office of Nuclear Physics under contract DE-AC05-06OR23177.

\appendix
\label{APPLFWF}
\section{Pion light front wavefunction}
On the light front, the pion state in the instanton vacuum in leading order in diluteness or packing fraction, is given by~\cite{Liu:2023fpj}
\begin{widetext}
\bea
\label{PWF0}
|P\rangle =
\int \frac{dx}{\sqrt{2x\bar x}}\int \frac{d^2k_\perp}{(2\pi)^3}\sum_{A,s_1, s_2}
\Phi_{\pi^a}(P, x,k_\perp;s_1,s_2)\,b_{s_1}^{A\dagger}(k)d_{s_2}^{A\dagger}(P-k)|0\rangle
\eea
with  
\bea
\label{PWF}
    \Phi_{\pi^a}(P, x,k_\perp;s_1,s_2)
    =\frac{1}{\sqrt{N_c}}\bigg[\phi_\pi(x, k_\perp)=\,
    \frac{C_\pi\mathcal{F}(\frac{k_\perp}{\lambda_\pi \sqrt{x\bar x}})}{{\sqrt{2x\bar{x}}}\big(m^2_\pi-\frac{k^2_\perp+M^2}{x\bar{x}}\big)}\bigg]
    \,\bar{u}_{s_1}(k)i\gamma^5\tau^a v_{s_2}(P-k)
    \eea
    \end{widetext}
with  on-shell free massive spinors, e.g. $(\slashed{k}-M)u(k)=0$.
(\ref{PWF0}) is obtained by diagonalizing the 
light front Hamiltonian in leading order in the
instanton diluteness or packing fraction.
It is an eigenstate of the light front squared mass 
$$2P^+P^-|P\rangle =(P_\perp^2+m_\pi^2)\,|P\rangle$$ 
 subject to  the light front normalization $\langle P|P\rangle=(2P^+)(2\pi)^3 \delta^3(0_P)\equiv 2P^+V^-$. Here $C_\pi$ is fixed by the normalization of the wave functions and $\lambda_\pi$ is fixed by the normalization of the pion distributions. Practically, $\lambda_\pi$ is chosen to be $2.464$ \cite{Liu:2023fpj}.
 The emergent fermionic form factor $\cal F$ is the profile of the instanton zero mode in singular gauge, 
 \bea
 {\cal F}(k)=\big(z(I_0(z)K_0(z)-I_1(z)K_1(z))'\big)^2|_{z=\frac 12 k\rho}\nonumber\\
 \eea
(\ref{PWF}) is composed of two correlated constitutive quarks  and no constitutive gluon. Hence the first
gluonic contribution in (\ref{PWF}) vanishes.

\section{Pion scalar form factor}
\label{APPSCALAR}

\subsection{Chiral Analysis}
The contribution to the pion scalar form factor is fixed by 
the strictures of broken chiral symmetry, which is upheld by the dilute QCD instanton vacuum.
With this in mind, the on-shell chiral Ward
identities give~\cite{Yamagishi:1995kr}
\begin{widetext}
\bea
\label{SCALAR}
F_S(q)= &&-\frac 1{f_\pi}
+\frac 1{f_\pi^2}\langle \sigma(0)\rangle\nonumber\\
&&-\frac{m_\pi^2}{f_\pi}
\int d^4x\,e^{iq\cdot x}
\langle T^*(\sigma(x)\sigma(0))\rangle_c\nonumber\\
&&-\frac 1{f_\pi^2} p_1^\alpha p_2^\beta\int d^4y_1d^4y_2\, e^{-ip_1\cdot y_1+ip_2\cdot y_2}
\langle T^*({\bf j}_{A\alpha}^a(y_1)
{\bf j}_{A\beta}^a(y_2)\sigma(0))\rangle_c
\eea
\end{widetext}
with  the scalar interpolating field identified as
$$m\overline\psi\psi(x)=-f_\pi m_\pi^2(f_\pi+\sigma(x))$$
and the 1-pion reduced axial current ${\bf j}_A$. The deviation from
the GOR relation  
$\langle\sigma (0)\rangle$  is relatively small.
 Space-like the contribution of the 2-pion treshold
at LO is 
 \bea
 \label{FS1}
 F_S(q)\approx -\frac 1{f_\pi}+\frac 1{f_\pi^3}
 \bigg(q^2+\frac 12 m_\pi^2\bigg)\,
\overline {\cal J}(q)
\eea
with $q^2=-Q^2$, and
\bea
\label{FS2}
\overline{\cal J}(q)=&&\frac 1{16\pi^2}\bigg(1+\frac{1}{x^2}\bigg)^{\frac 12}
{\rm ln}\bigg(\frac{\sqrt{1+x^2}-x}{\sqrt{1+x^2}+x}\bigg)\nonumber\\
&&+\frac{1}{16\pi^2}\left(2+\frac{17}5\frac{q^2}{q^2+\frac12 m_\pi^2}\right)
\eea
with $x^2=Q^2/4m_\pi^2$ following a minimal subtraction \cite{Bijnens:1998fm,Bijnens:1997vq}.
%We note that
%\bea
%F_S(q)=-\frac 1{f_\pi}
%\bigg(1+\frac{q^2}6\langle r^2\rangle_S+{\cal O}(q^4)\bigg)
%\eea
The  chiral contribution to the pion sigma term  form factor, follows from
\bea
\sigma_\pi(q)=\frac{\langle p_2|m\overline\psi\psi(0)|p_1\rangle}{2m_\pi^2}=-\frac 12 f_\pi F_S(q)
%\frac 12-\frac 1{2f_\pi^2}
 %\bigg(q^2+\frac 12 m_\pi^2\bigg)\,
%\overline{\cal J}(q)
\nonumber\\
\eea
after using  (\ref{FS1}) to one-loop,  
in agreement with (\ref{FHX}) in the forward limit.

\subsection{Sigma meson FF}
The pion sigma term form factor in the ILM can be obtained
in the mean-field approximation, by resumming the t-channel bubble diagrams between the constitutive quarks induced by the emergen 't Hooft interactions. 
The result in the zero size approximation is
\begin{equation}
    \sigma_{\pi}(Q^2)=\frac{\sigma_\pi(0)}{1+Q^2/m_{\sigma}^2}
\end{equation}
where $m_{\sigma}=683$ MeV \cite{Liu:2024}
and $\sigma_\pi(0)\approx \frac 12$.

\bibliography{PION}

\end{document}